\documentclass[12pt]{article}

\usepackage{graphics}
\usepackage{amssymb}
\usepackage{theorem}
\usepackage{graphicx}   
\usepackage{latexsym}
\usepackage{psfrag}
\usepackage{color}
\usepackage{amsmath}
\usepackage{natbib}
\usepackage{fancybox}
\newcommand{\balpha}{{\mbox{\boldmath $\alpha$}}} 
\newcommand{\btheta}{{\mbox{\boldmath $\theta$}}} 
\newcommand{\bbeta}{{\mbox{\boldmath $\beta$}}}

\newcommand{\bx}{{\mbox{\boldmath $x$}}} 
 
\newcommand{\bsigma}{{\mbox{\boldmath $\sigma$}}} 
 
\newcommand{\bean}{\begin{eqnarray*}}
\newcommand{\eean}{\end{eqnarray*}}

\newcommand{\ignore}[1]{}

\definecolor{darkgreen}{rgb}{0.2,.7,0.2}
\definecolor{darkblue}{rgb}{0.2,0.2,0.7}
\definecolor{orange}{rgb}{.9,0.3,0}

\begin{document}
\title{Bayesian radiocarbon modelling\\for beginners}
\author{Caitlin E.\ Buck \& Miguel Juarez}
\date{University of Sheffield}

 \maketitle

  \tableofcontents
  
 \begin{abstract}
 Due to freely available, tailored software, Bayesian statistics is fast becoming the dominant paradigm in archaeological chronology construction. Such software  provides users with powerful tools for Bayesian inference for chronological models with little need to undertake formal study of statistical modelling or computer programming. This runs the risk that it is reduced to the status of a black-box which is not sensible given the power and complexity of the modelling tools it implements. In this paper we seek to offer intuitive insight to ensure that readers from the archaeological research community who use Bayesian chronological modelling software will be better able to make well educated choices about the tools and techniques they adopt.  Our hope is that they will then be both better informed about their own research designs and better prepared to offer constructively critical assessments of the modelling undertaken by others.
 \end{abstract}

\section{Background}

  Bayesian chronological models are statistical models that allow us to represent, manage and interpret both relative and absolute chronological information from one or more archaeological or palaeoenvironmental research projects.  They were developed over the last thirty years specifically for the archaeo and palaeo research communities by statisticians and software developers who took advantage of a revolution in our ability to implement such models using a simulation-based (as opposed to an exact calculation) approach.

  Users of the resulting software need not know the details of the underlying maths and stats nor of the computational techniques used to implement them.  They must, however, understand the concept of a model, appreciate the choices they are making when they select a particular model to represent their own project and understand enough of the key decisions made by the statistical modellers and software developers to know which software is appropriate for their needs. 
  
  In this paper, we aim to provide readers with some of the background they need to undertake each of these tasks.  In Section~\ref{sec.good_model} we look at some key concepts and decisions involved in modelling in general and then Section~\ref{sec.stat_chrono} focusses on the basics of formal chronology construction, highlighting the fact that section drawings and Harris Matrices are both types of chronological model. Formal statistical notation is then introduced, in Section~\ref{sec.c14}, and used to define the statistical models now in routine use for Bayesian radiocarbon dating. These are then applied to specific examples and illustrative software output is discussed.  Section~\ref{sec.implementation} focuses on some of the practicalities involved in using Bayesian radiocarbon calibration software, and Section~\ref{sec.future} looks to the future.

\subsection{What makes a good model?}
\label{sec.good_model}
For the purposes of this paper, we define a model as:
\begin{quote}
a representation of a person, organism, structure or concept typically smaller, simpler and/or more abstract than the original.
\end{quote}

This definition highlights the very many different ways in which the term model is used in modern parlance, but it tells us nothing about what makes a good model for any specific purpose. To move towards this, we start by thinking about a simple and perhaps rather trivial modelling problem: what makes a good model of an elephant?

\begin{figure}
\begin{center}
  \includegraphics[width=3in]{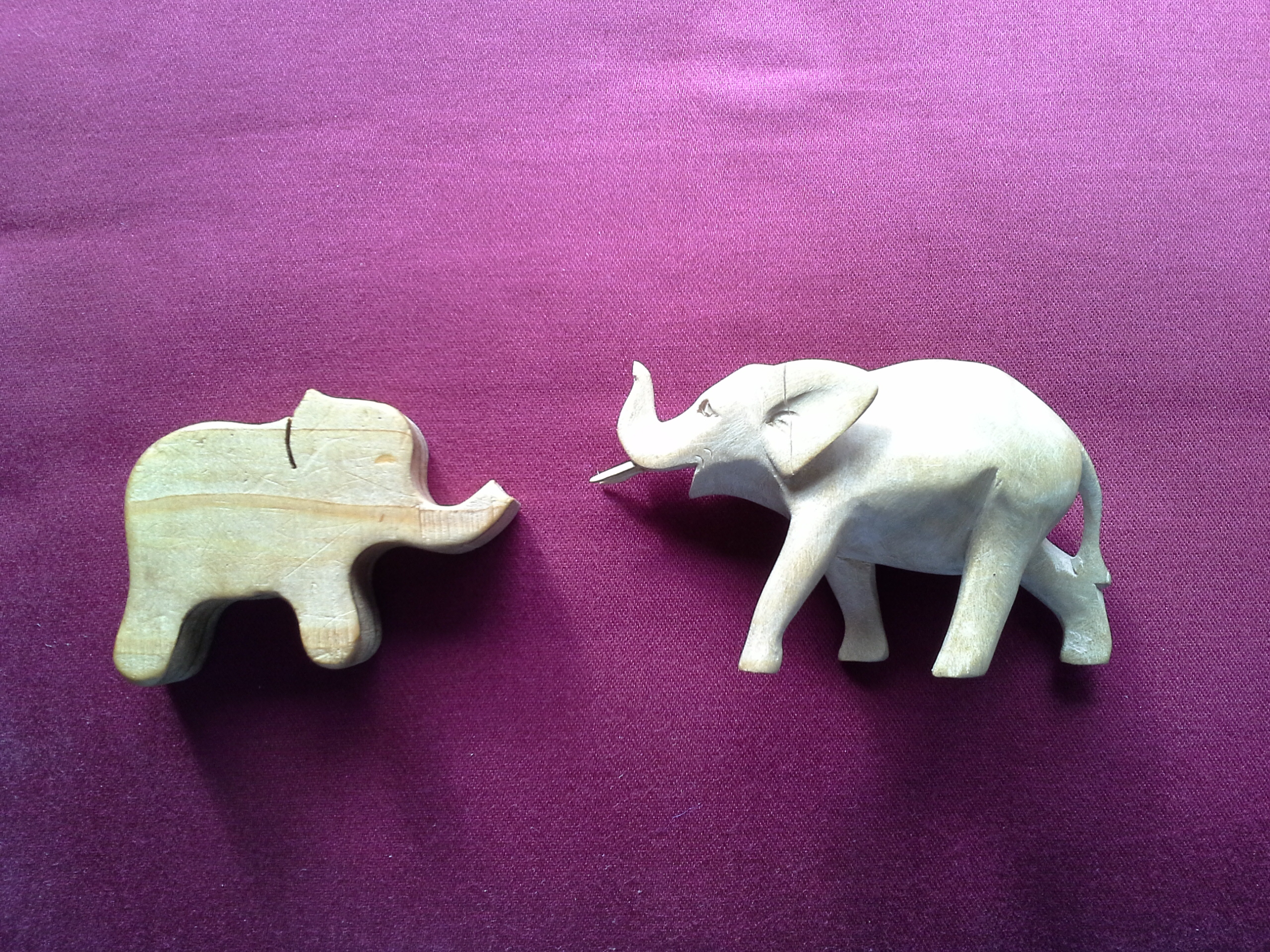}
 \caption{Two models of elephants: which is best?}
 \label{fig.elephants}
\end{center}
\end{figure}

Consider the model elephants pictured in Figure \ref{fig.elephants}; clearly neither are anatomically accurate, but is one model better than the other? Our contention is that which one is best depends on what the model is to be used for.  If the model is for entertaining a 3-year-old child on a wet afternoon, then the one on the left is probably the best.  It's safe for them to play with on their own, has no sharp or fragile pieces and will allow them to recognise an elephant just from its long nose. If, on the other hand, the model is to be used to help a 10-year-old child learn the key features of real elephants, then the model on the right is surely more suitable.  It has more realistic legs and head and also has a tail, ears and tusks which are all missing from the one on the left.  Were we to want to move beyond these basics, however, to teach an older child or adult about, say, the differences between African and Indian elephants or about the physiology of elephants relative to other large mammals then neither of the models pictured would be suitable and we would need to look elsewhere for something more anatomically detailed.

Analogies of this sort are useful only up to a point and this one could certainly be taken too far, but before we leave it and move to think about chronological models it is worth noting a couple of similarities between them and model elephants.  Both chronological models and model elephants can be: off-the-shelf or tailor-made and descriptive or mechanistic. The elephants in Figure~\ref{fig.elephants} are both off-the-shelf and descriptive, but were we to seek anatomically correct models for more sophisticated purposes then they may well need to be tailor-made and mechanistic. 

In a similar way, for many purposes, chronological models that are descriptive and available off-the-shelf can be all that is required to complete our archaeological or palaeoenvironmental inference.  In other situations we might need off-the-shelf, but mechanistic chronological models which capture (in part) the mechanisms that led to the chronological observations we have made.  In some situations, however, none of the models on offer in existing off-the-shelf software are suitable for our needs and then we must ask a statistician or software developer to construct a tailor-made, probably mechanistic, model for us.  

\section{Formal chronology construction}
\label{sec.stat_chrono}

All modelling clearly involves making choices.  Generally, we start by selecting a meduim, framework or paradigm in which to construct the model and other choices then follow.  In statistical modelling, our first choice is between the Frequentist or Classical paradigm and the Bayesian one. Frequentist/Classical statistics is based on a classical interpretation of probability which defines an event's probability as the limit of its relative frequency in a large number of trials.  Bayesian statistics --- named for Thomas Bayes (1701--1761) --- is a paradigm in which  evidence about the true (but unknown) state of the world is expressed in terms of `degrees of belief', represented as personal probability statements.  Because of it's focus on degrees of belief, the Bayesian paradigm is ideally suited to the representation and management of expert opinion and prior knowledge as well as scientific data and this makes it particularly appealing to archaeologists and palaeoenvironmental scientists looking for a coherent way to draw together information from several different sources.

Buck et al (1996) made the case for the use of the Bayesian paradigm in archaeology in some detail and we do not recap those arguments here.  Instead, we consider the circumstances we find ourselves in when constructing chronological models and propose that, for all but the simplest problems, the Bayesian paradigm seems most natural.

\subsection{Appropriate chronological models}

Just as with the elephant modelling problem, in order to choose the appropriate nature and scale of model for our archaeo or palaeo research project, we need to know precisely what we are modelling and why.  For example, if we are seeking to date a single event in the archaeological record, like the death of an individual human whose articulated skeleton has been found in a well sealed grave, then a Frequentist approach might be suitable. We could (in theory) repeatedly date the same event by sending multiple samples from the skeleton to the laboratory for dating and then summarise the results within the Classical statistical paradigm.  However, if we want to date a sequence of events (e.g.\ stratigraphy from an archaeological site or a sediment core) then the relative chronological knowledge needs modelling and, probably, some expert chronological knowledge/opinion too. In which case we will surely need the Bayesian paradigm. If we want to construct a chronology for a whole site or landscape then multiple sequences need comparing and combining: several experts might be involved, working at different times and/or in different places and so the model must be modular and will almost certainly need to account for personal probability statements from multiple experts.  In such circumstances, the Bayesian paradigm is the only one that we are aware of that allows robust interpretation of all of the inter-related sources of information simultaneously.

\subsection{Pictures as chronological models}
 
\begin{figure}
\begin{center}
 \includegraphics[width=3in,trim={5cm 26cm 5cm 5cm}, clip]{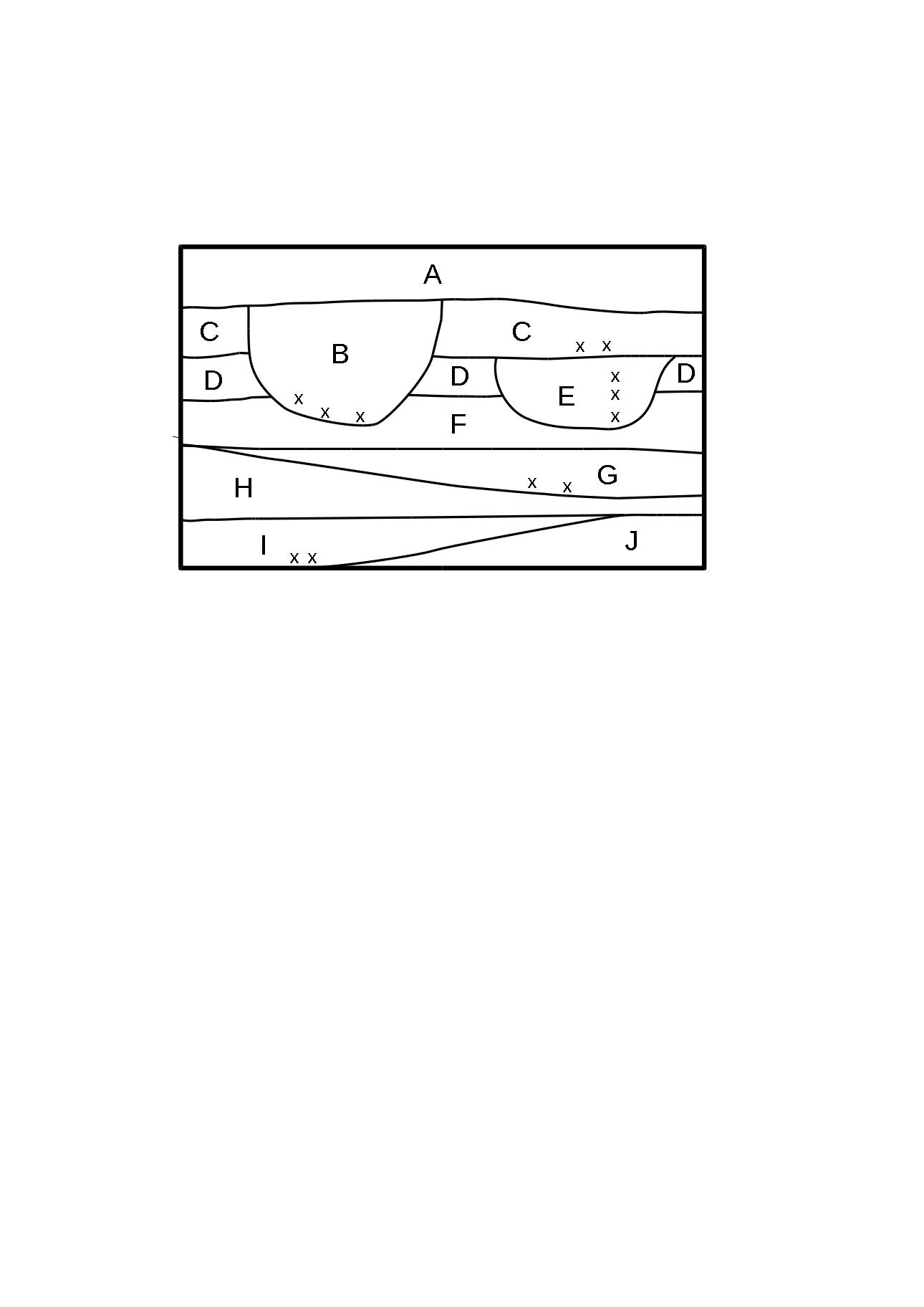}
\end{center}
\caption{An illustrative section drawing from part of an (imagined) archaeological site. The locations of samples suitable for chromometric dating are indicated by crosses.}
\label{fig.strat}
\end{figure}

Almost all chronologists who are drawn towards the Bayesian paradigm are motivated in part by the desire to combine scientific dates and relative or absolute prior knowledge. Stratigraphic information, is the most common such knowledge from archaeological excavations and is usually held in the research archive in the form of sketches or plots: section drawings, phase diagrams, Harris matrices or age/depth plots.  These are all pictorial models of relative chronology and are a key starting point in chronological modelling.  Figure~\ref{fig.strat} offers an example section drawing from an imagined archaeological site which we use to illustrate how such information can be utilised as the starting point for chronological modelling.

\begin{figure}
\begin{center}
\begin{tabular}{c c c c}
 \raisebox{.4in}{\includegraphics[width=.3in]{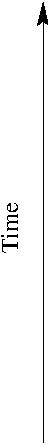}}&\includegraphics[trim={5cm 4cm 10cm 2cm},clip,width=2in]{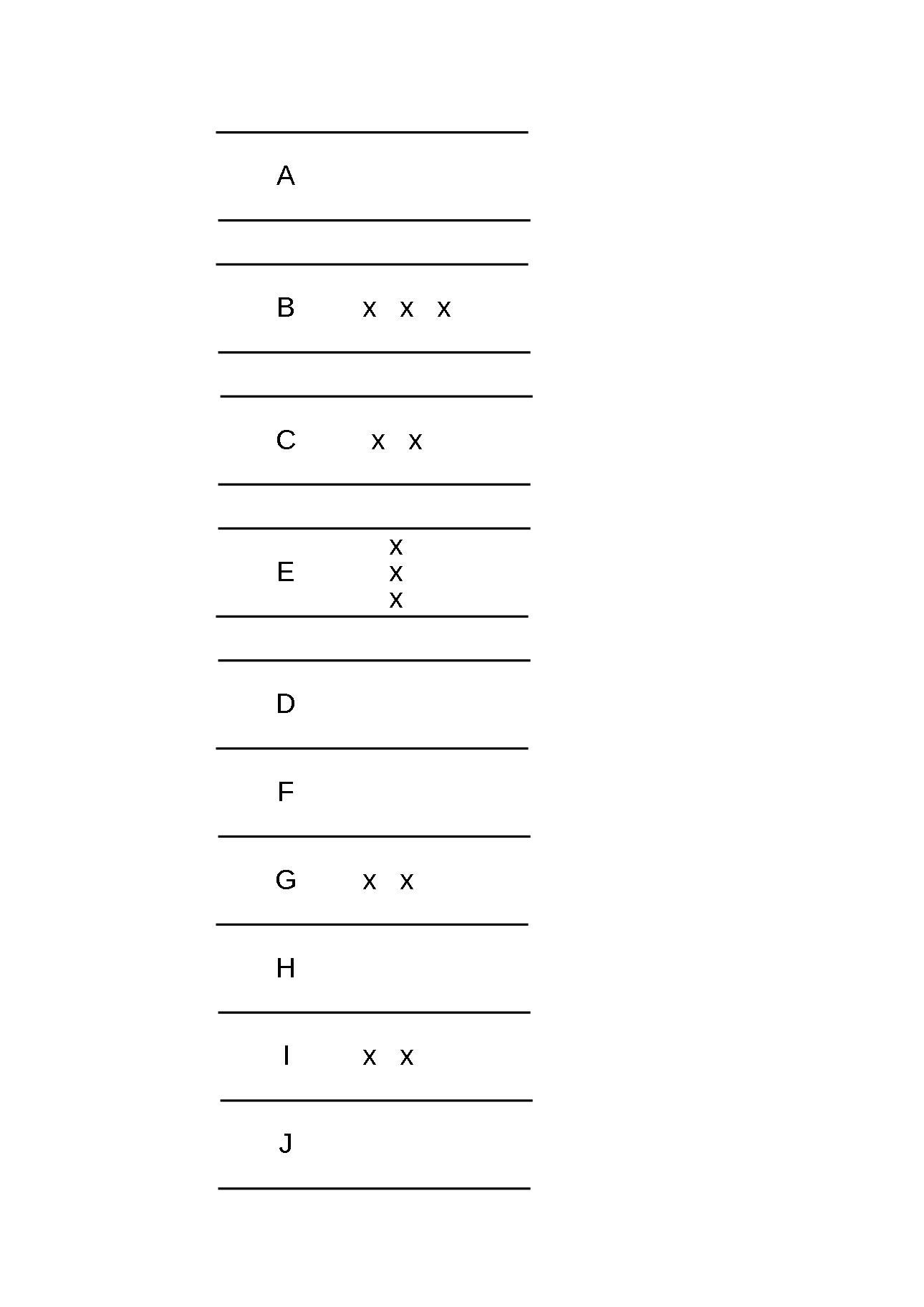}&\raisebox{1.5in}{$~~~~~ $or$~~~~~$}&\includegraphics[trim={5cm 4cm 10cm 2cm},clip,width=2in]{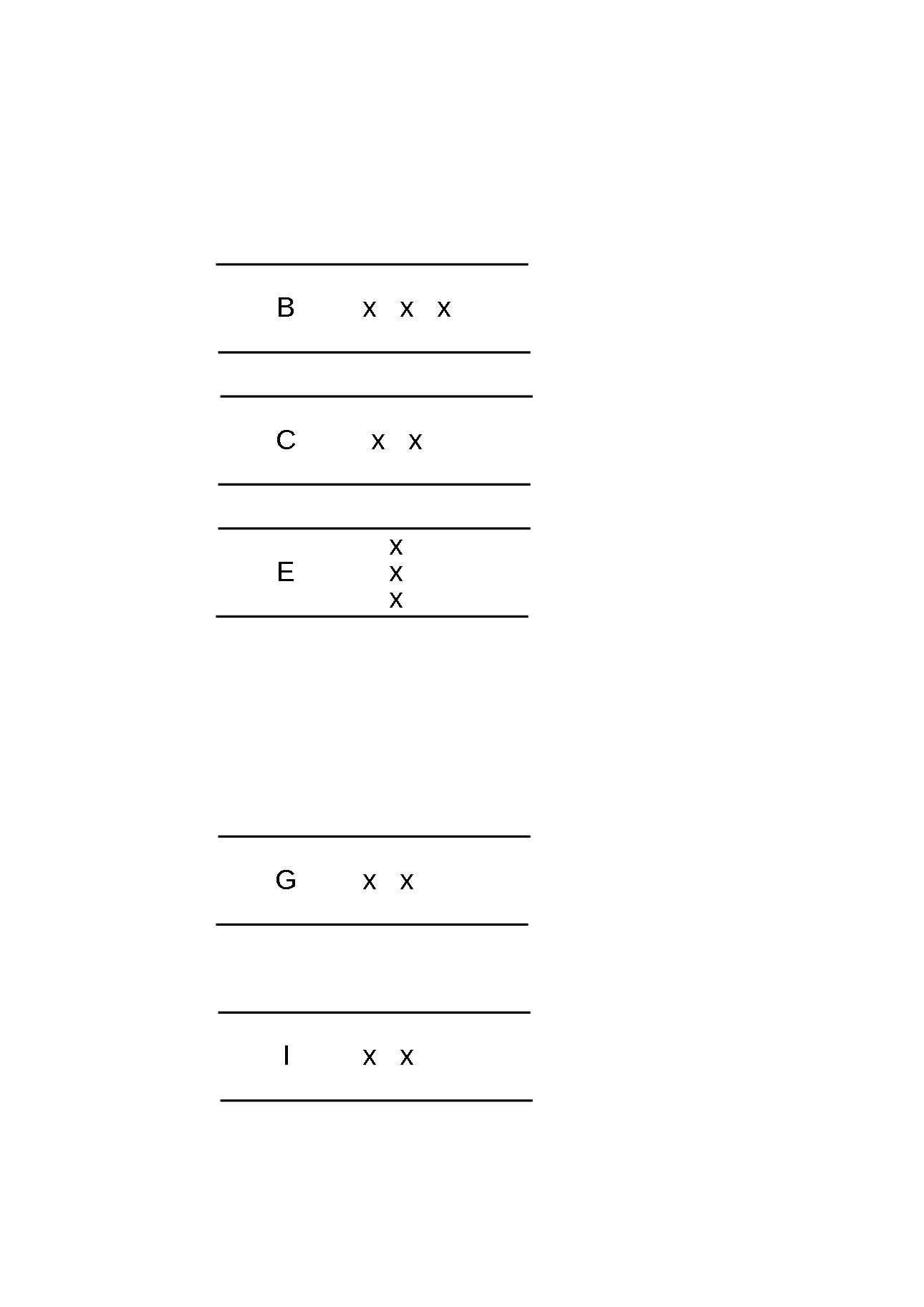}
\end{tabular}
\end{center}
\caption{Sketches of the relative chronological information contained within the stratigraphic profile shown in Figure~\ref{fig.strat}. Horizontal lines represent archaeological context boundaries and crosses indicate the location within the stratigraphic sequence of samples suitable for chronometric dating. The profile on the left shows all of the contexts in Figure~\ref{fig.strat}; the one on the right shows only the contexts that contain samples suitable for absolute dating.}
\label{fig.model}
\end{figure}

The first step in creating a formal chronological model is to simplify the stratigraphic drawing and to focus purely on the temporal information it contains.  The left-hand sketch in Figure~\ref{fig.model} shows just such a simplification for the illustrative section in Figure~\ref{fig.strat}.  The right-hand sketch in the same figure shows a simplification of the one on the left, with a focus only on contexts that contain samples that could be submitted for chronometric dating. 

By drawing the two sketches in Figure~\ref{fig.model} side-by-side, we highlight a key choice that all chronological modellers must make i.e.\ which contexts and samples to include. Most modellers are agreed that, all other things being equal, we should follow Occam's razor and keep the model as simple as possible.  This seems like good, straightforward advice, but in practice all other things are seldom equal. By which we mean that excluding or including contexts or samples from our chronological model, and precisely how we represent the ones we do include, will almost always have at least some impact on the results we get.

Given this, and the fact that there are many more such choices we need to make as we undertake the implementation process, in the later sections of this paper we offer some general guidance for those seeking to make responsible use of Bayesian chronological modelling software. Before we can do that, however, we need to a) clearly identify the key chronological components that we wish to manage or interpret and b) to think and write rather more formally, thus constructing statistical models.   We address a) in the next section and move to b) in Section~\ref{sec.c14}. In doing so we draw on a large body of existing literature, but in particular: \citet{Naylor.Smith-1988, Buck.Cavanagh.Litton-1996, Blackwell.Buck-2008} and citations therein.

\subsection{Key components of a statistical chronological model} 

There are broadly two types of information to be represented in a statistical model for chronology construction: relative and absolute.  Relative chronological information typically relates to the (prior) ordering of events; whereas absolute chronological information usually arises from historical records or from scientific dating methods.  

In the remainder of this paper, we focus on absolute dates that arise from radiocarbon dating, but Bayesian methods have also been developed for dendrochronology \citep{Litton.Zainodin-1991,Millard-2002,Jones-2013}, luminescence dating \citep{Zink-2015} and electron spin resonance dating \citep{Millard-2006}.  We will also centre what we say around seeking to manage and interpret the chronological information represented in Figures~\ref{fig.strat} and \ref{fig.model}.  Bayesian models now exist to represent a considerably wider range of chronological features than those needed for this purpose, but our goal here is to be introductory rather than comprehensive and we hope that interested readers will read further work by the authors cited herein.

There are two key types of chronological event in Figures~\ref{fig.strat} and \ref{fig.model}: those that relate to directly datable objects (like the deposition of the samples indicted by crosses) and those, like the creation of context boundaries, that do not.  Absolute date estimates for the context boundaries can only be obtained by modelling their relationship with the datable objects, via the stratigraphic sequence.  So, in summary we need model components to represent the following:
\begin{itemize}
 \item the true underlying dates we wish to learn about, only some of which relate directly to datable objects,
 \item stratigraphic relationships between the true underlying dates of all components of the stratigraphic record,
 \item the relationship between the true underlying dates and radiocarbon determinations, including laboratory uncertainties and the necessary calibration.
\end{itemize}
In the next section we look at all three of these, starting with a simple model which includes only directly datable objects and their associated stratigraphic relationships and then moving to include context boundaries and their stratigraphic relationships to the datable objects. In so doing, we suppose that the datable samples in Figures~\ref{fig.strat} and \ref{fig.model} (identified by crosses) give rise to the radiocarbon determinations indicated in Table~\ref{tbl.c14}.

\begin{table}
\begin{scriptsize}
\begin{center}
 \begin{tabular}{|c |c |c| c| c|}
 \hline
Context	&Sample label	&Stratified within context&Mean C14 age	&Lab error\\
\hline
B	&$\theta_1$&No	&5700	&30\\
	&$\theta_2$&No	&5670	&30\\
	&$\theta_3$&No	&5650	&30\\
\hline
C	&$\theta_4$&No	&5720	&30\\
	&$\theta_5$&No	&5780	&30\\
\hline
E	&$\theta_6$&Yes	&5900	&50\\
	&$\theta_7$&Yes	&5870	&50\\
	&$\theta_8$&Yes	&5850	&50\\
\hline
G	&$\theta_9$&No	&6000	&30\\
	&$\theta_{10}$&No	&6130	&30\\
\hline
I	&$\theta_{11}$&No	&6200	&50\\
	&$\theta_{12}$&No	&6250	&50\\
\hline
 \end{tabular}
 \end{center}
 \end{scriptsize}
\caption{Radiocarbon determinations assumed to be associated with the illustrative stratigraphic sequence in Figure~\ref{fig.strat}, along with an indication as to whether or not we are assuming each sample to be stratified within the relevant context.}
\label{tbl.c14}
\end{table}

\section{Models for Bayesian radiocarbon dating}
\label{sec.c14}

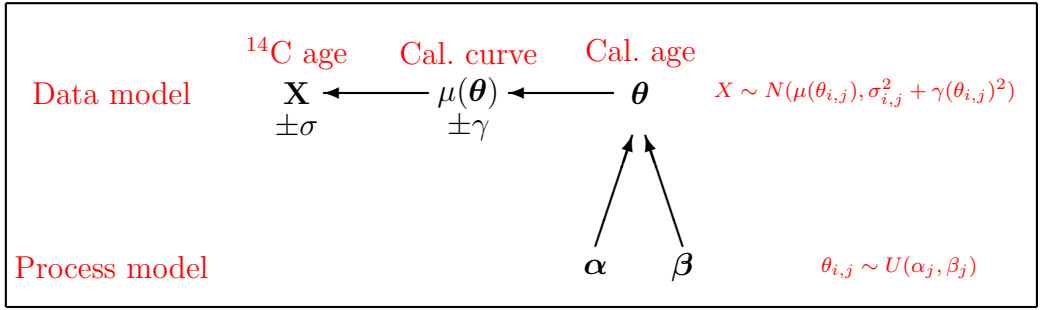
\begin{figure}
$~~~~~~~~~~~~~~~$
\begin{picture}(500,50)  
\thicklines
\put(0,15){\makebox(0,0){\textcolor{red}{Data model}}}
\put(70,30){\makebox(0,0){\textcolor{red}{$^{14}$C age}}}
\put(200,30){\makebox(0,0){\textcolor{red}{Cal.\ age}}}
\put(135,30){\makebox(0,0){\textcolor{red}{Cal.\ curve}}}

\put(135,15){\makebox(0,0){$\mathbf{\mu}(\mathbf{\btheta})$}}
\put(70,15){\makebox(0,0){$\mathbf{X}$}}
\put(200,15){\makebox(0,0){$\mathbf{\btheta}$}}

\put(120,15){\vector(-1,0){40}}
\put(190,15){\vector(-1,0){40}}

\put(70,2){\makebox(0,0){$\pm \mathbf{\sigma}$}}
\put(135,2){\makebox(0,0){$\pm \mathbf{\gamma}$}}

\put(285,15){\makebox(0,0){\scriptsize{\textcolor{red}{$X \sim N(\mu(\theta_{i,j}),\sigma_{i,j}^2+\gamma(\theta_{i,j})^2)$}}}}
\end{picture}

$~~~~~~~~~~~~~~~$
\begin{picture}(500,50)
\thicklines
\put(0,0){\makebox(0,0){\textcolor{red}{Process model}}}
\put(183,0){\makebox(0,0){$\mathbf{\balpha}$}}
\put(183,8){\vector(1,3){14}}
\put(216,0){\makebox(0,0){$\mathbf{\bbeta}$}}
\put(216,8){\vector(-1,3){14}}
\put(298,0){\makebox(0,0){\textcolor{red}{\scriptsize{$\theta_{i,j} \sim U(\alpha_{j},\beta_{j})$}}}}

\put(-40,-15){\line(0,1){115}}
\put(-40,-15){\line(1,0){390}}
\put(350,-15){\line(0,1){115}}
\put(-40,100){\line(1,0){390}}
\end{picture}
$~~~$
\vspace{0.25in}
\caption{Pictorial representation of the (hierarchical) statistical model developed by \citet{Naylor.Smith-1988}.  Here we represent the relationship between chronological parameters that relate directly to data (i.e.\ $\btheta$, $\bx$ and $\bsigma$) and those used to represent the underlying archaeological processes ($\balpha$ and $\bbeta$).  To emphasize the process relationships, the arrows represent the causal direction which is the opposite direction from the one in which we make inferences.}
\label{fig.graphical}
\end{figure}

Since the focus in this section is on formal statistical modelling, some readers may find it daunting.  For those who do, we suggest that you focus on appreciating the notation used and on the general structure of the equations provided.  It is not essential to understand the details of the equations to gain insight into the structure of the models and the general nature of the way in which they are constructed and it is these that are the most important. Given this, we offer Figure~\ref{fig.graphical} which we hope readers will use to follow the structure of the model as it is described.  The top part of the figure relates to the ideas in Sections~\ref{sec.basic} and the lower part to those in Section~\ref{sec.deposition}.

Before we get into details, however, it is worth highlighting a few notational conventions. We use Greek and Roman letters to represent individual components of statistical models which are known as parameters.  Some parameters take single numerical values (known as scalars), and we indicate these using standard font.  Parameters defined using a bold font indicate a fixed length sequence of scalars, known as a vector.  Since vectors contain a sequence of values, we often undertake calculations systematically for each entry in the vector.  For example, if $\mathbf{z}$ is a vector of length $k$ (i.e.\ contains a sequence of $k$ scalars), then to multiply the elements in $\mathbf{z}$ together, we would write $\prod_{m=1}^k z_m$, where $\prod$ indicates multiply and $z_m$ (a scalar) is just one of the entries of $\mathbf{z}$.

We will also need to use the term probability distribution function (or probability density) which refers to a function whose value at any given point provides a probabilistic statement about the parameter represented by the function.  The familiar bell-shaped curve of the Normal probability distribution function is one that readers will probably be able to call to mind.  Such distributions can be used one-at-a-time to represent individual, independently varying parameters.  More commonly in statistical models, however, we work with collections of such functions that are inter-related or covarying.  When we do this, we typically undertake calculations on lots of variables simultaneously (or jointly) so that we can keep track of their inter-relationships.  Then, when we create plots of the final results, we usually focus on variables one at a time since visualising high dimensional distributions is notoriously difficult.  Such single variable representations are referred to as marginal (as opposed to joint) probability distribution functions.

\subsection{The basic Bayesian radiocarbon model}
\label{sec.basic}

A radiocarbon determination has two parts: a radiocarbon age estimate, $x$ before present (BP), and an associated laboratory error, $\sigma$.   We want to use this determination to learn about the (true underlying) calendar date, $\theta$ calibrated BP (cal BP), on which the sample sent for dating ceased metabolising. To help us we often have prior information (historical, stratigraphic, etc.) about the calendar date, which we represent with the (prior) probability density $p(\theta)$.

In order to use the data and the prior information to learn about the calendar date, we need to formalize the link between the information we have,  i.e.\ $x\pm\sigma$ and $p(\theta)$, and what we want to learn about i.e.\ $p(\theta| x, \sigma)$ (note that the symbol~$|$ can be read as `given' so this is the probability density of $\theta$ given $x$ and $\sigma$). Since the proportion of radioactive carbon atoms in the earth's atmosphere has not been constant over time, we need a calibration curve to map between radiocarbon and calendar ages. We refer to this calibration curve as $\mu(\theta)$.  Hence, strictly, we want to learn about $p(\theta| x, \sigma, \mu(\theta))$.

Since we would not get precisely the same radiocarbon determination if we sent several parts of the same organic sample to the same radiocarbon laboratory, $x$ BP is just one realisation of a random variable $X$ BP, which is associated with a specific calendar date $\theta$ cal BP. We can think of $X$ as the true value of the calibration curve at date $\theta$ plus some uncertainty, $\epsilon$.  If the true value of the calibration curve at $\theta$ cal BP is $\mu(\theta)$ BP then 
$
X=\mu(\theta) + \epsilon
$
and, since $\epsilon$ is usually assumed to be Normally distributed, $\epsilon \sim N(0,\sigma^2)$.

Focussing on context E from Figure~\ref{fig.strat}, for example, we have three radiocarbon determinations from a stratigraphic sequence (see Table~\ref{tbl.c14}) and so $\bx = (5900, 5870, 5850)$ and $\bsigma = (50, 50, 50)$ and the true underlying dates associated with them are $\btheta=(\theta_6,\theta_7,\theta_8)$. The stratigraphic information allows us to be sure (a priori) that the three calendar dates are strictly ordered, $\theta_6 > \theta_7 > \theta_8$. Then, using the standard statistical model for radiocarbon \citep[not motivated here, but discussed in detail in][Chap.\ 9]{Buck.Cavanagh.Litton-1996}, 
$$
P(\bx|\btheta) \propto \prod_{i=m}^n \exp \left\{ -\frac{ (x_i - \mu(\theta_i))^2}{2\sigma_i^2+2 \gamma(\theta_i)^2} \right \}
$$
where $\gamma(\theta_i)$ is the standard deviation on the current internationally-agreed estimate of the calibration curve at $\theta_i$ and, in our example $m=6$ and $n=8$. Note that in the formulation in \citet{Naylor.Smith-1988} and \citet{Buck.Cavanagh.Litton-1996} the parameter $\gamma(\theta_i)$ is not present since those authors assumed this to be very small relative to $\sigma_i$ which was reasonable at the time, but no longer holds because laboratory errors have reduced considerably.

The (prior) stratigraphic information can then be formalised as
\[
P(\btheta) \propto I(\theta_i) = \begin{cases}
    1  & \theta_{i-1} > \theta_i > \theta_{i+1} \\
    0 & \text{otherwise}
  \end{cases}
\]
and the Bayesian solution (or posterior distribution function) is 
\begin{equation}
P(\btheta|\bx) \propto P(\bx|\btheta) \times P(\btheta) \propto \left\{ \prod_{i=m}^n \exp \left\{-\frac{(x_i - \mu(\theta_i))^2}{2 \sigma_i^2+2 \gamma_{\mu(\theta_i)}^2} \right \} \right \} I(\btheta) .
\label{eqn.Bayes}
\end{equation}

\begin{figure}
\begin{center}
\begin{center}
\includegraphics[width=5in]{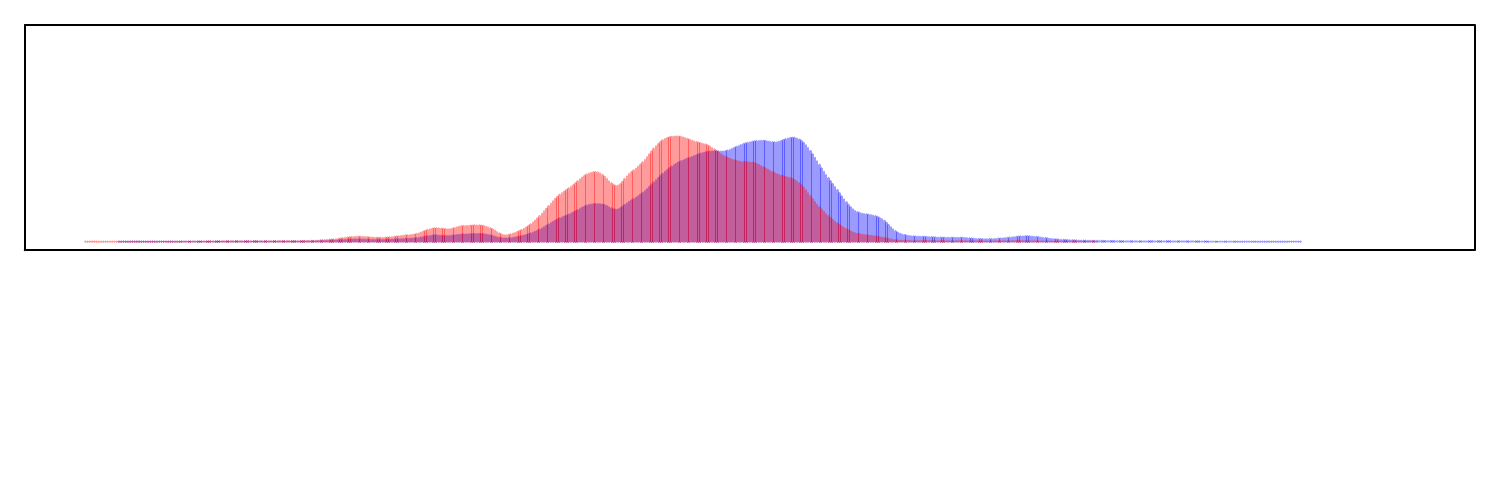}
\vspace{-.75in} \\
\includegraphics[width=5in]{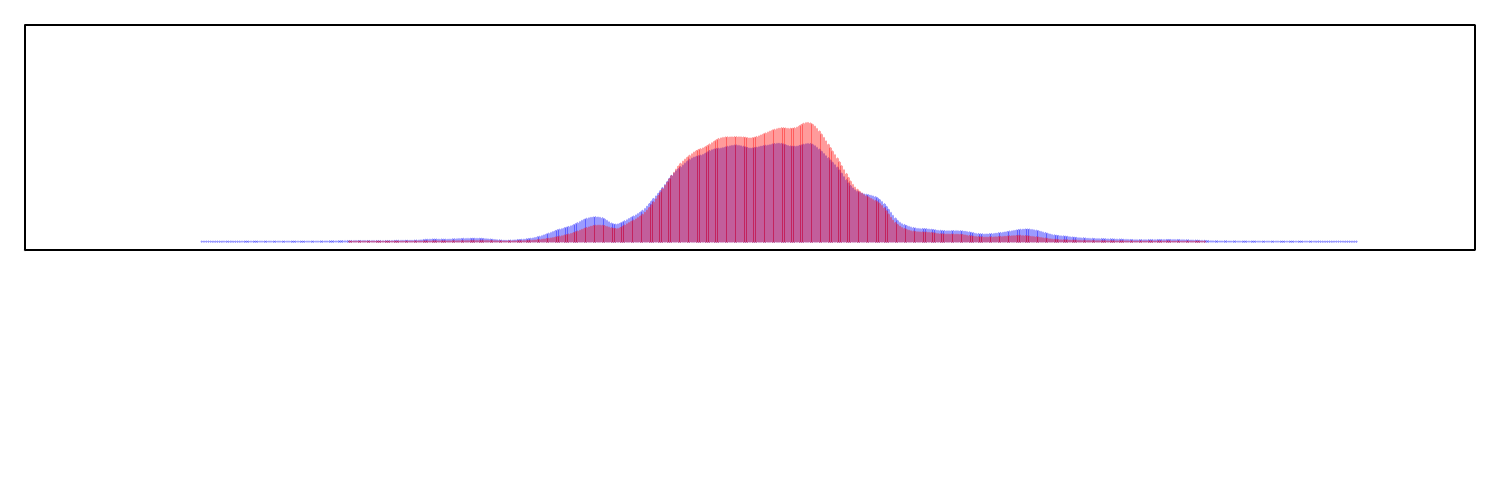}  
\vspace{-.75in} \\
\includegraphics[width=5in]{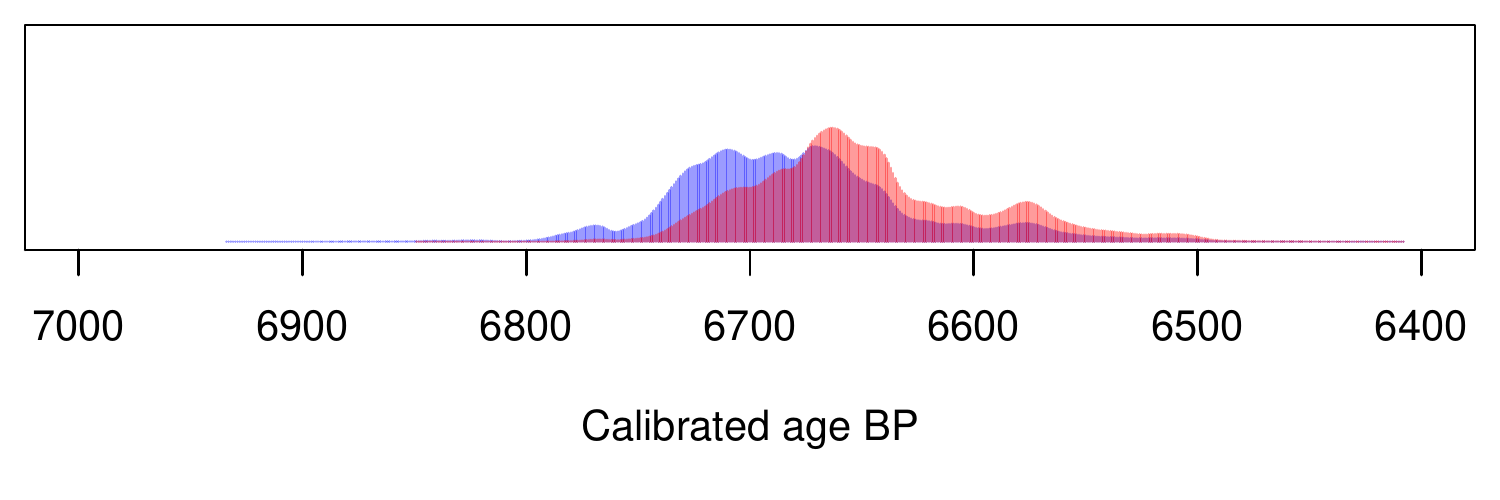}
\end{center}
\caption{Results of calibrating the three radiocarbon determinations 5900$\pm$50 BP, 5870$\pm$50 BP and 5850$\pm$50 BP (blue when stratigraphic info is ignored, red when it is included by application of Equation~\ref{eqn.Bayes}).}
\label{fig.simple}
\end{center}
\end{figure}

With modern computational power and simulation-based implementation methods \citep[detailed in][and discussed below]{Buck.Cavanagh.Litton-1996}, applying this equation to even very large stratigraphic sequences is now straightforward using off-the-shelf Bayesian radiocarbon calibration packages such as OxCal \citep{OxCal-BR-2009} or BCal \citep{BCal}. For illustration, we used BCal with the IntCal13, internationally-agreed, estimate of the radiocarbon calibration curve \citep{Reimer.etal-2013} to calibrate the three stratified determinations from context E in Figure~\ref{fig.strat}  and obtained the results shown in Figure~\ref{fig.simple}, where blue indicates the probability distribution function obtained when each determination is calibrated individually and red indicates the results from computing the Bayesian posterior density function in Equation~\ref{eqn.Bayes}.  

This simple Bayesian model is clearly powerful, allowing inclusion of basic stratigraphic information that would otherwise have been ignored or handled in an ad hoc manner.  In so doing, we typically obtain more precise date estimates than those that can be obtained by calibrating determinations individually.  For context E, the gain in precision is fairly modest because we have only three determinations that are temporally rather widely spaced.  With more data in a similar time interval, however, the gains can be considerably more substantial. Despite these benefits, the model in Equation~\ref{eqn.Bayes}  does not have sufficient complexity to allow us to include all of the features of the stratigraphy in Figures~\ref{fig.strat} or \ref{fig.model} since there is no way to represent events for which we do not have direct dating evidence.  In particular we have no way to represent the concept of depositional history, as recorded by the relationships between archaeological contexts.

\subsection{Modelling deposition}
\label{sec.deposition}

To represent boundaries between different depositional contexts or phases, \citet{Naylor.Smith-1988} introduced two further parameters
to the model, $\balpha$ and $\bbeta$ (both cal BP) with $\alpha_j>\beta_j$ (i.e.\ $\alpha_j$ and $\beta_j$ are, respectively, the early and late boundary dates for context $j$). Typically, we have some relative chronological information, which relates such parameters to datable material, but there is no direct scientific dating evidence associated with them. This situation is very common, of course, since in archaeology and palaeoenvironmental research we seldom find datable material directly associated with all of the key locations in our stratigraphic sequences.  

Naylor and Smith (1988) thus amended the model in Equation~\ref{eqn.Bayes} so that for contexts with no internal stratigraphy (and hence no a priori ordering of the dates of the samples)
$$
P(\bx|\btheta) \propto
\prod_{j=1}^J \left\{ 
(\alpha_j-\beta_j)^{-n_j}
\prod_{i=1}^{n_j} z_{i,j}
I_B(\theta_{i,j})
\right\}
$$
where $J$ is the number of contexts or phases in the model, $n_j$ is the number of datable samples in context or phase $j$,
$$
z_{i,j}= \exp \left\{
-\frac{(x_{i,j} - \mu(\theta_{i,j}))^2}{2\sigma_{i,j}^2+2 \gamma(\theta_{i,j})^2}\right \}, 
$$
and
\[
I_B(\theta_{i,j}) = \begin{cases}
    1  & \beta_j \ge \theta_{i,j} \ge \alpha_j\\
    0 & \text{otherwise} .
  \end{cases}
\]

Assuming that the deposition rate for material in each context is constant, but allowing varying deposition rates between contexts, they then modelled the prior knowledge as
\[
P(\btheta) \propto I_A(\balpha, \bbeta) = \begin{cases}
    1  & \balpha, \bbeta  \in A \\
    0 & \text{otherwise}
  \end{cases}
\]
where $A$ is the set of values of $\balpha$ and $\bbeta$ that satisfy the prior chronological (e.g.\ stratigraphic) information.  Thus obtaining calibrated dates via
\begin{equation}
P(\btheta|\bx) \propto I_A(\balpha, \bbeta) \times \prod_{j=1}^J \left\{ 
(\alpha_j-\beta_j)^{-n_j}
\prod_{i=1}^{n_j} z_{i,j}
I_B(\theta_{i,j})
\right\}.
\label{eqn.phase}
\end{equation}

In situations where we have stratigraphic ordering of samples within a context (as in context E), it is computationally straightforward to add the relevant parts of Equation~\ref{eqn.Bayes} when we implement Equation~\ref{eqn.phase} in software.  It is also trivial to allow for contexts or phases for which there is stratigraphic information, but no direct dating evidence at all (i.e.\ pairs of ${\alpha_j, \beta_j}$ that have direct relation only to other boundary parameters and not to any samples that can be directly dated).

\begin{figure}
\begin{center}
\includegraphics[width=5in]{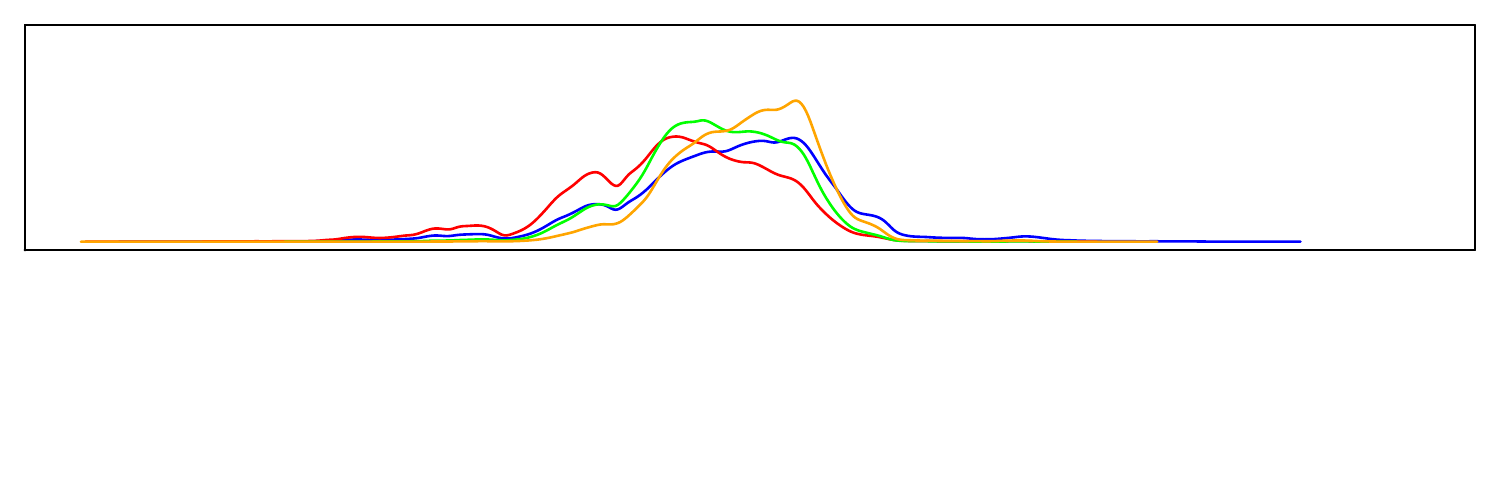} 
\vspace{-.75in} \\
\includegraphics[width=5in]{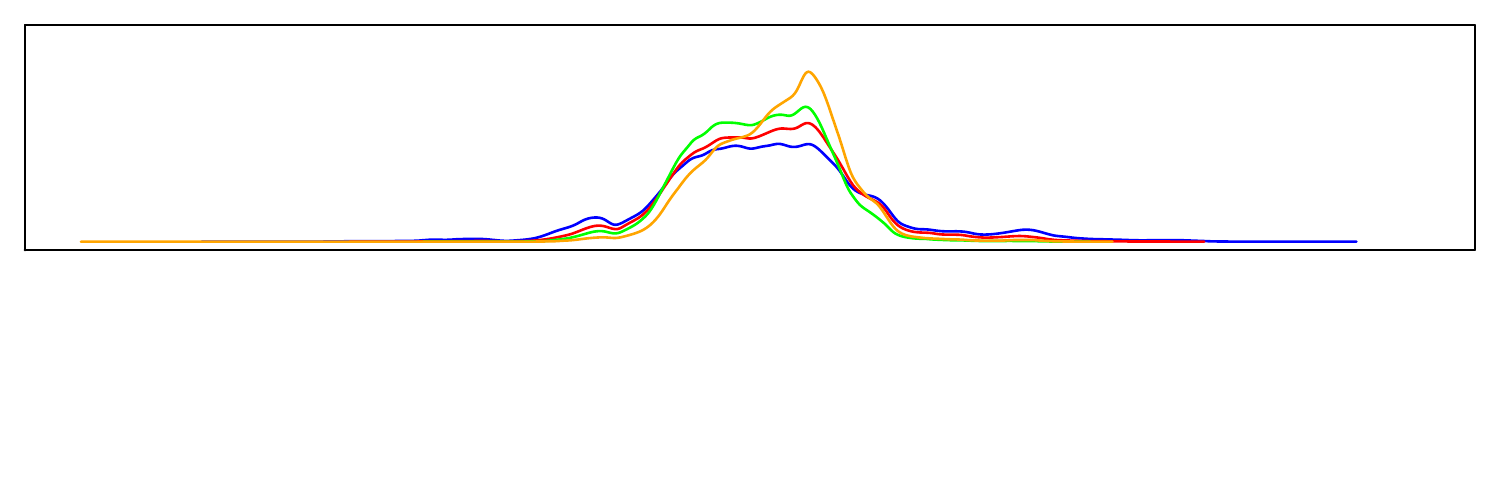} 
\vspace{-.75in} \\
\includegraphics[width=5in]{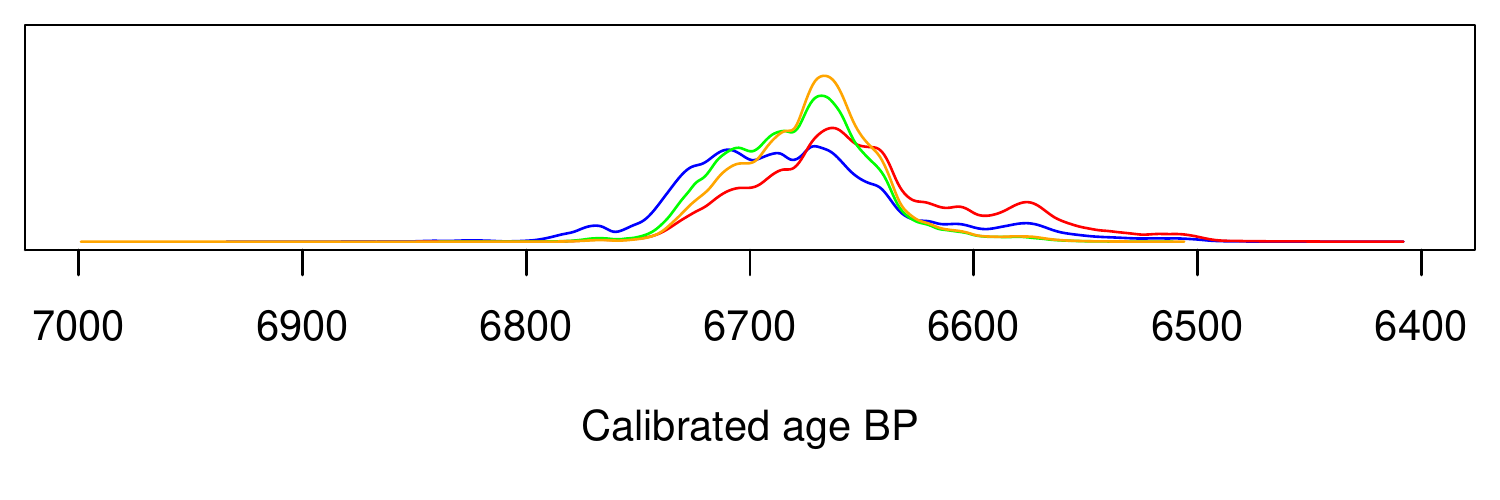}
\end{center}
\caption{
Results of calibrating the three radiocarbon determinations represented in context E. \textbf{Blue} when
all stratigraphic info is ignored; \textbf{red} when ordering ony within this context is included (using Equation~\ref{eqn.Bayes}); \textbf{green} when ordering within this context is included \textbf{and} the contextual relationships shown in the right-hand panel of Figure~\ref{fig.model} are modelled (using Equation~\ref{eqn.phase}); \textbf{orange} when ordering within this context is included \textbf{and} the contextual relationships shown in the left-hand panel of Figure~\ref{fig.model} are modelled (using Equation~\ref{eqn.phase}). Note a) that we are showing only the outline of the marginal probability density in each case so that all four distributions  can clearly be seen and b) that as we add more stratigraphic detail to our model the probability densities become increasingly peaked (i.e.\ the date estimates are more precise).
} 
\label{fig.full_plot}
\end{figure}

OxCal \citep{OxCal-BR-2009} and BCal \citep{BCal} offer tools to implement the full range of such models and we adopt the latter to formalise the models in Figure~\ref{fig.model}.  When we do this we are, of course, embedding context E within a larger model for the whole of the stratigraphic sequence shown in Figure~\ref{fig.strat}. Nonetheless, for illustration, we can focus just on the marginal results for the dates of the samples within that context, thus producing Figure~\ref{fig.full_plot}, which suggests that as we add more stratigraphic detail (and hence more parameters) to our model we (appear to) learn increasingly precisely about the dates for the samples in context E.

Superficially, this seems attractive since greater precision is almost always the goal of chronologists, however we need to be cautious here.  Surely there must be a point at which adding more contexts to our model, without adding any more absolute dating evidence leads to a false sense of extra information.   Does knowing that there are an extra two contexts between contexts G and E really provide very much extra chronological information (given that neither are directly associated with any absolute dating evidence) and, if so, is our statistical modelling of the information in Figure~\ref{fig.graphical} capturing it well? This and other implementation questions are discussed in the next section but before we move to that, we look first at some of the estimates obtained for the dates of the context boundaries in our example. 

\begin{figure}
\begin{center}
\includegraphics[width=5in]{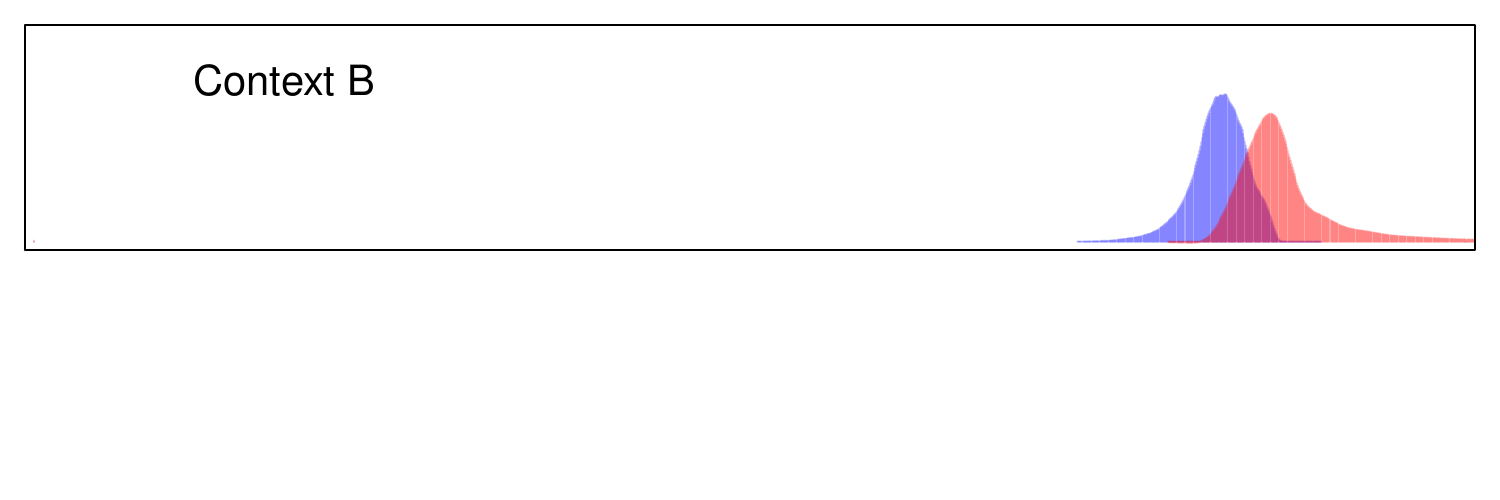} 
\vspace{-.75in} \\
\includegraphics[width=5in]{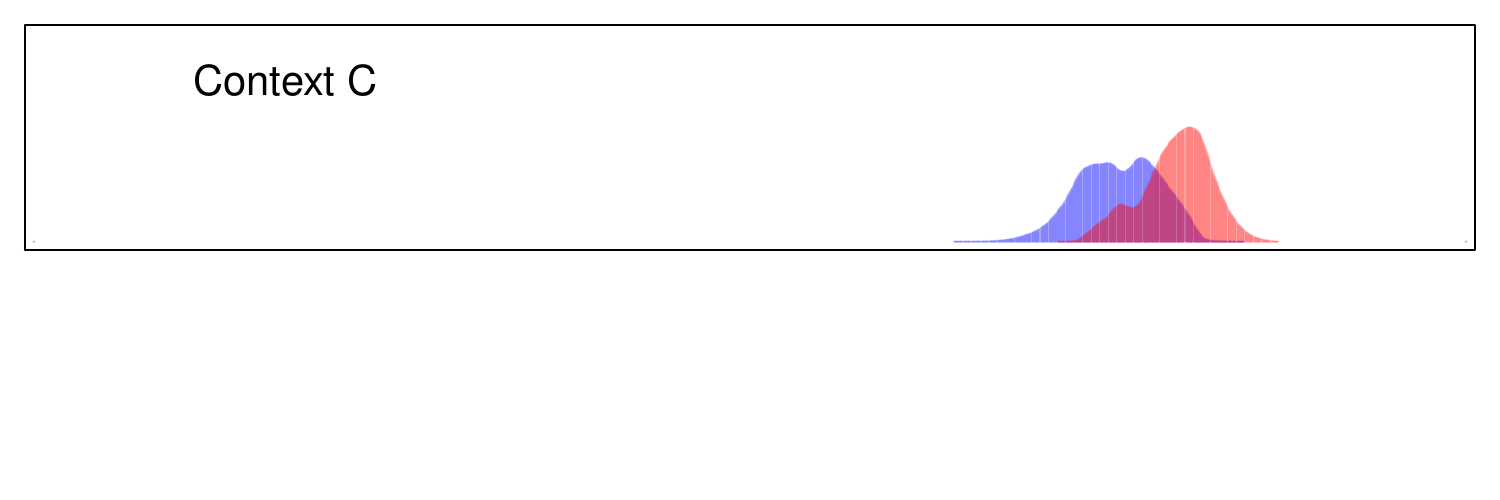} 
\vspace{-.75in} \\
\includegraphics[width=5in]{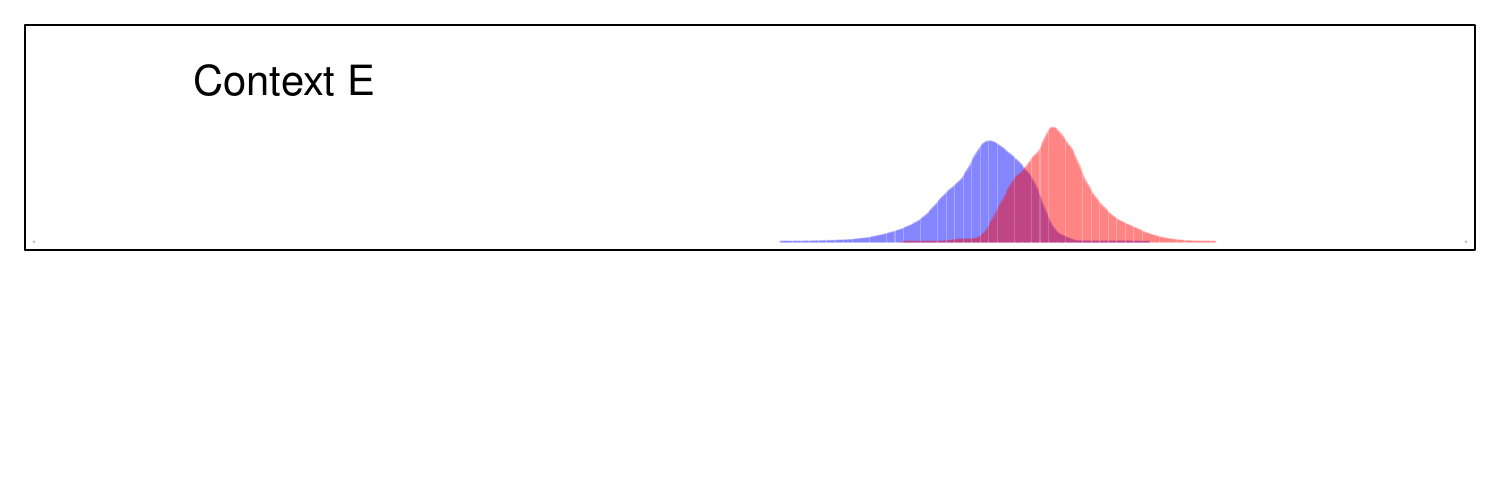} 
\vspace{-.75in} \\
\includegraphics[width=5in]{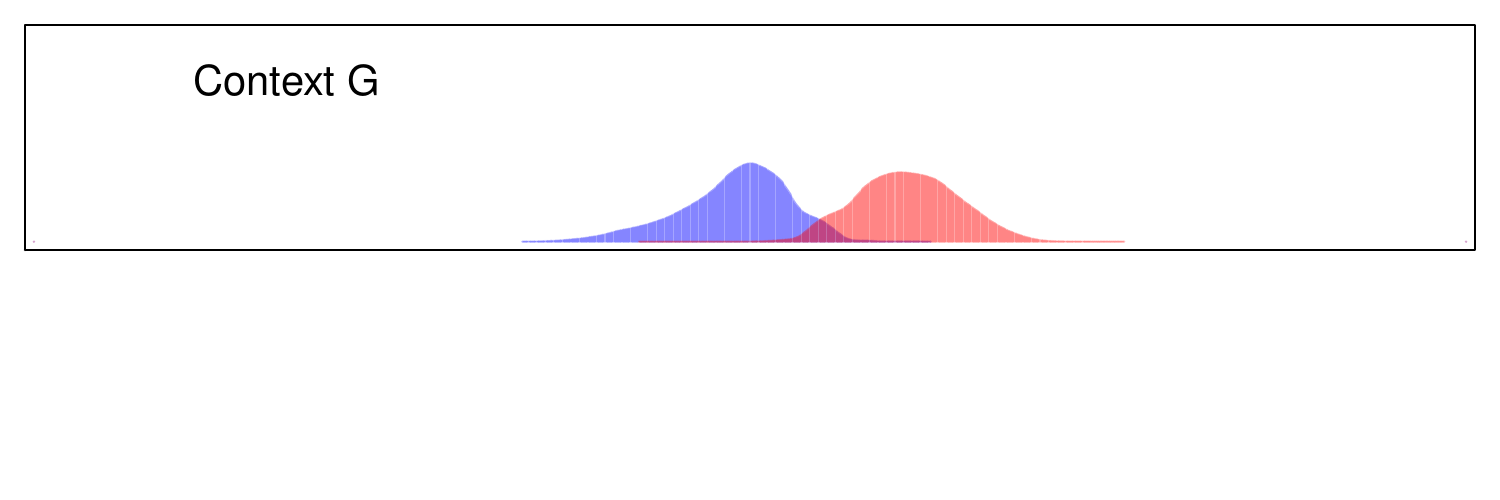}  
\vspace{-.75in} \\
\includegraphics[width=5in]{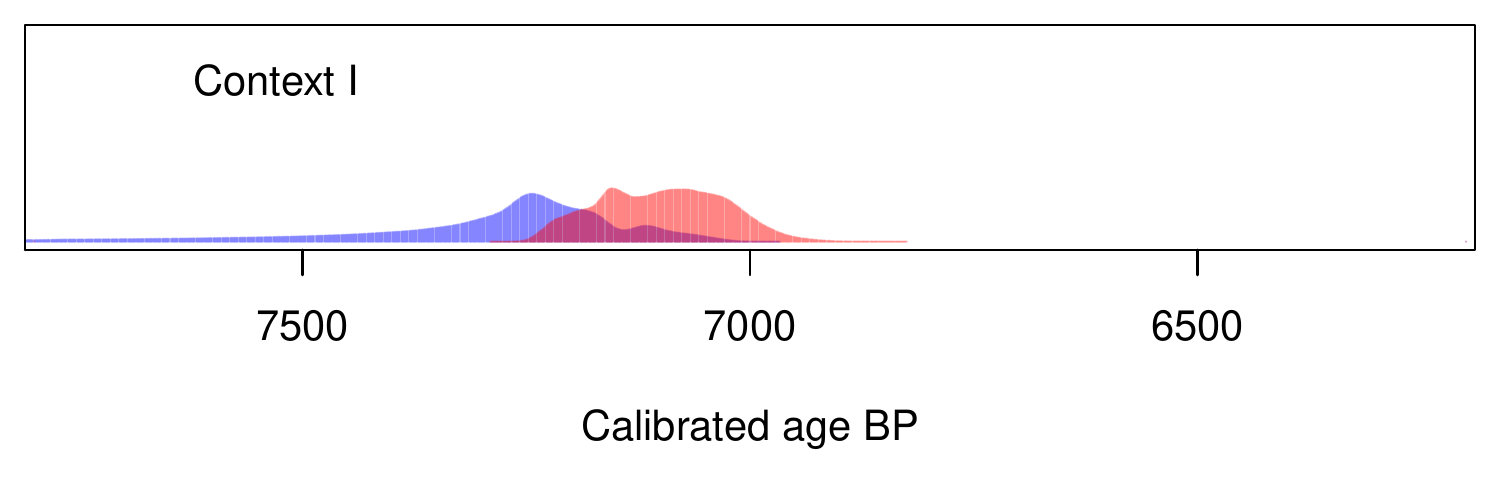}  
\end{center}
\caption{Estimated calibrated context boundary dates for the contexts shown in the righ-hand panel of Figure~\ref{fig.model}. Early boundary daets for each context are in blue and late ones in red.} 
\label{fig.boundaries}
\end{figure}

Given standard model selection advice (summarised above and discussed further below), all-other-things being equal we prefer simple over more complex models and so Figure~\ref{fig.boundaries} shows the posterior estimates of the boundary dates for the contexts in the right-hand sequence of Figure~\ref{fig.model}. Similar results could be shown for the left-hand sequence, should there be a strong archaeological reason for doing so.  What we see in Figure~\ref{fig.boundaries} is the clear impact of the temporal ordering imposed by the prior distribution on the boundary parameters, which were derived from the context relationships shown in Figure~\ref{fig.model}.  We also find that we can date each of the context boundaries to within at most 500 years and in the case of context B, to within 100 years, despite the fact that we have no  direct absolute dating evidence for any of these parameters. Clearly this is a very powerful archaeological inference tool,  even for this very simple stratigraphic sequence.  Before using such methods to aid in the interpretation of real archaeological projects, however, we must return to some of the inferential and modelling decisions related to their use, particularly the modelling dilemma outlined in the previous paragraph.

\section{Modelling and implementation decisions}
\label{sec.implementation}

\subsection{Model selection}

When \citet{Naylor.Smith-1988} first proposed the models in Equations~\ref{eqn.Bayes} and \ref{eqn.phase} they implemented them using a technique known as numerical integration which is time consuming to programme and the resulting code is computationally demanding.  Taken along with the fact that at-the-time computational power was extremely limited, this meant that only the simplest of archaeological models could be implemented.  The rapid increase in adoption of Bayesian chronological modelling did not come, therefore, until several years later when coders started to adopt simulation-based methods for implementation, known as Markov chain Monte Carlo (MCMC) methods.  These are considerably easier to code initially and also to modify when new components are needed, thus making them more attractive for modular modelling problems of the sort discussed here and highly amenable to the implementation of generalisable algorithms like those used in OxCal and BCal. 

With modern computational power and these faster, more flexible implementation tools a range of software options are now available to chronological modellers.  This has led to an ever increasing choice of modelling options and selecting precisely which model to implement is thus a routine dilemma.  Currently, the only widely used formal model choice tools for chronologists are those offered by OxCal \citep{OxCal-BR-2009}. Put simply, these compare the prior with the posterior probability densities and prefer models that result in posterior densities that are closest to the priors.  

Such methods are somewhat arbitrary, however, and best suited to comparing models of similar structure and size and so are not ideal for addressing the dilemma we faced at the end of Section~\ref{sec.deposition}.  For that we really need a formal model selection technique and these are notoriously difficult to construct, especially when the numbers of parameters vary greatly between the various representations, as can easily be the case with large archaeological sequences.  For this reason model selection is an art that Bayesian chronological modellers need to acquire and which, like most arts, can only really be learnt by experience. 

The present authors have some experience of such modelling and, in general with Occam's razor in mind, we prefer models with fewer parameters over those with more.  So, in the case of the dilemma in the previous section, we preferred the model represented by the right-hand panel of Figure~\ref{fig.graphical}, over that on the  left.  We do this because the simpler model allows representation of the key depositional history for the datable samples, but does not explicitly include those contexts that contain no absolute dating evidence and so add relatively little extra information to the inference process.

We might quite readily make a different choice however, if the archaeological event that we most needed to date related to one of the contexts that contains no datable samples.  For illustration, suppose that context F were a destruction layer between two periods of human activity, then we may well decide to include context F in the model, accepting that the total amount of chronological information we have available has not increased very much, but allowing us to do our best to date the context about which we are most eager to learn.

\subsection{Responsible use of software}

Given what we have said about implementation of chronological modelling software, coding it is clearly not a task for a novice and so we assume that readers of this paper will be using one of the off-the-shelf packages like OxCal or BCal.  These are very powerful tools when used carefully, but they do come with some responsibilities.  These relate to gaining at least an intuitive insight into what models the software is implementing and how.  This paper provides a basic introduction to some aspects, but users should also read the specific background literature cited by the software providers and the explanations offered by the user manuals.

We all also have responsibilities when writing up modelling work.  It is vital, for example, that readers are not simply told which model (or software) was used, but that they are also offered a clear explanation as to why the model in question was selected for the current application.  Typically, as in the examples above, some reasons will relate to the author's interpretation of the archaeology and some to theoretical or practical constraints.  Such details are important because scientific dating evidence is often collected before models are specified and so the only way that readers can be sure that the same sources of information were not used multiple times is for all authors to offer clear justification for their choices of model and prior and that these do not rely on any aspect the scientific dating evidence.

Users should also know enough about how the software implements their model that they can make principled decisions and report clearly on them. This is important because MCMC-based software uses simulation and so each time it is run (slightly) different results are obtained. Key concepts here include:
\begin{itemize}
\item Burn-in: samples discarded from the beginning of MCMC simulations, which may not be very representative of the posterior density. This is necessary because all simulations start from somewhat arbitrarily selected values which might not have high posterior probability.
\item Convergence: the desired state of the MCMC sampler, from which samples can reliably be used to estimate the posterior.  The burn-in phase is intended to ensure that the samples we store are from this part of the simulation.
\item Thinning and effective sample size: the process of storing only a sub-set of MCMC samples (thinning) to leave a subset which is smaller than the total but nonetheless conveys equivalent information (and so has the same effective sample size). Subsetting the output in this way is most important when parameters of our model are highly correlated (as is often the case in stratigraphic sequences) and so the simulated values for each parameter change only very slowly between neighbouring steps of the Markov chain.
\end{itemize}
Most software providers, offer automated convergence checking tools that provide guidance about the amount of burn-in and thinning required. Howerver, users should understand intuitively the checks being conducted and record the choices made in their write-up so that others may replicate and/or build appropriately on their work. The OxCal and BCal manuals both offer more advice on  these issues.

Since MCMC methods are simulation-based, users must check reproducibility by making multiple runs for each model, with different start values for the sampling chains, to check that the results obtained for key parameters are the same to an \textit{appropriate} level of accuracy.  \textit{Appropriate} is, of course, a relative term. We will require a different level of accuracy in historic periods from the palaeolithic.  Reproducibility experiments, and the accuracy to which we report results, should
reflect this.

Since applied Bayesian statistics involves so many personal judgements users should also explore the sensitivity of results to the choices they made.  For each application users should vary models and/or priors, to explore other plausible options, and report resulting changes in the posteriors densities, just as we did here when we considered two formalisations of the stratigraphic information in Figure~\ref{fig.strat}. Such checks are essential since without them we have no idea how robust results are.  If they aren't robust to key decisions, which for large or complex models is not unusual, considerable further exploration may be needed as to why particular choices were made.

\section{Other current and future options}
\label{sec.future}

There are a considerable number of modelling options now available to users of off-the-shelf Bayesian chronological modelling software and we will not attempt to discuss here all, or even very many, of these.  The following are worth highlighting, however:
\begin{itemize}
\item model extensions to allow inclusion of a wide range of absolute prior knowledge not discussed here. For example: that derived from historical documents or classical texts and/or from scientific dating evidence other than radiocarbon.
\item a range of deposition models e.g.\ of the gradual colonisation of a site \citep{Jones-2013, Lee.BR-2012} and of peat or sediment accumulation \citep{Christen.Clymo.Litton-1995, Haslett.Parnell-2008, BR-2008, Blaauw.Christen-2005, Blaauw.Christen-2011}.
\item detection of  outliers in radiocarbon dating \citep{Christen-1994, BR-2008}.
\item automated selection of samples during an incremental radiocarbon dating programme in order to use the dating budget cost effectively \citep{Christen.Buck-1998,Buck.Christen-1998}.
\end{itemize}
 The first three of these are in routine use and are implemented in several freely available software packages, in particular OxCal \citet{OxCal-BR-2009}.  The fourth is not however and this seems rather strange given the very large dating projects now undertaken within the Bayesian framework.  Of course, uptake is likely to be limited until such methods are offered in freely available software and so it would be good to see them developed in the near future.  OxCal already has a feature, known as R\_Simulate, which allows users to generate likely additional radiocarbon dates for potential samples within an existing model. This is the first step towards such a tool, but at present the automatic selection of samples likely to lead to the most cost effective dating programme is not.
 
Other desirable features not currently available in any of the standard software are 
\begin{itemize}
 \item extensions of the purely temporal Bayesian chronological models to include spatial components, thus creating a spatio-temporal modelling framework.  Such extensions are desirable, in projects relating to the spread of humans or ideas in time and space.  However, they are computationally very demanding to implement since they require data from entire landscapes to be analysed simultaneously.  As a result there are two commonly adopted approaches to such problems.  The simplest is to spatially partition the data and then analyse the spatial groups within a purely temporal model \citep[as, for example, in][]{Blackwell.Buck-2003}.  A more sophisticated approach is to develop a mechanistic (typically deterministic) model for the spatial process (such as demographic spread) and then to use statistical methods to compare the resulting spread patterns with the available chronological evidence using formal statistical methodology \citep[as, for example, in][]{ Baggaley.etal-2012b, Baggaley.etal-2012a}.
 \item tools to automate production of pictures like those in Figure~\ref{fig.graphical}, from archaeological site databases and then to generate from them chronological models of the sort explored herein.  \citet{Dye.Buck-2015} provide a proof of concept about how such a tool might be developed, using a refinement of Harris matrices and techniques adopted from graph theory, but there is considerable work to do before such approaches would be ready for routine use.
 \item improvements to age-depth models to allow for geomorphology.  At present, such models effectively assume that the cores from lake sediments were derived from cylindrical lake basins. In practise we do not know the morphology of most of the lakes from which cores are taken, but we can be fairly sure that they are not cylindrical and recent work by \citet{Bennett.Buck-2016} shows that basin geomorphology can have a considerable impact on the age-depth relationships.  Given this, more work is needed to a) find a good, cost-effective ways to establish basin morphology and b) develop the existing Bayesian age-depth models to take account of the information obtained.
\end{itemize}

Clearly, such extensions would take chronology construction considerably beyond the relatively simple, but powerful models and methods outlined above.  Some may eventually be fairly readily added to the general purpose chronological modelling software like OxCal and BCal, but others, in particular the spatio-temporal modelling options, require inclusion of parameters with completely different structure from those used thus far.   As a consequence, at least for the moment, those wishing to adopt such models will need to learn not just modelling stills, but computer programming too.  The chronological modelling community is short of such skills and we hope that one or two readers of this paper might already have programming skills and be keen to help or might be willing to learn them  in order to do so.  We would be delighted to hear from anyone interested in such work and to encourage them to help us to make the next thirty years of chronological modelling as productive as the last.

\bibliography{biblio}

\begin{thebibliography}{}

\bibitem[Baggaley et~al., 2012a]{Baggaley.etal-2012b}
Baggaley, A.~W., Boys, R.~J., Golightly, A., Sarson, G.~R., and Shukurov, A.
  (2012a).
\newblock {Inference for population dynamics in the Neolithic period}.
\newblock {\em Annals of Applied Statistics}, 6:1352--1376.

\bibitem[Baggaley et~al., 2012b]{Baggaley.etal-2012a}
Baggaley, A.~W., Sarson, G.~R., Shukurov, A., Boys, R.~J., and Golightly, A.
  (2012b).
\newblock {Bayesian inference for a wavefront model of the Neolthisation of
  Europe}.
\newblock {\em Physical Review E}, 86.

\bibitem[Bennett and Buck, 2016]{Bennett.Buck-2016}
Bennett, K.~D. and Buck, C.~E. (2016).
\newblock Interpretation of lake sediment accumulation rates.
\newblock {\em Holocene}, 26(7):1092--1102.

\bibitem[Blaauw and Christen, 2011]{Blaauw.Christen-2011}
Blaauw, M. and Christen, J. (2011).
\newblock Flexible paleoclimate age-depth models using an autoregressive gamma
  process.
\newblock {\em Bayesian Analysis}, 6:457--474.

\bibitem[Blaauw and Christen, 2005]{Blaauw.Christen-2005}
Blaauw, M. and Christen, J.~A. (2005).
\newblock Radiocarbon peat chronologies and environmental change.
\newblock {\em Journal of the Royal Statistical Society: Series C (Applied
  Statistics)}, 54(4):805--816.

\bibitem[Blackwell and Buck, 2003]{Blackwell.Buck-2003}
Blackwell, P.~G. and Buck, C.~E. (2003).
\newblock {The Late Glacial human reoccupation of north western Europe: new
  approaches to space-time modelling}.
\newblock {\em Antiquity}, 77(296):232--240.

\bibitem[Blackwell and Buck, 2008]{Blackwell.Buck-2008}
Blackwell, P.~G. and Buck, C.~E. (2008).
\newblock Estimating radiocarbon calibration curves.
\newblock {\em Bayesian Analysis}, 3:225--248.

\bibitem[Bronk~Ramsey, 2008]{BR-2008}
Bronk~Ramsey, C. (2008).
\newblock Deposition models for chronological records.
\newblock {\em Quaternary Science Reviews}, 27(1--2):42--60.

\bibitem[Bronk~Ramsey, 2009]{OxCal-BR-2009}
Bronk~Ramsey, C. (2009).
\newblock Bayesian analysis of radiocarbon dates.
\newblock {\em Radiocarbon}, 51(1):337--360.

\bibitem[Buck et~al., 1996]{Buck.Cavanagh.Litton-1996}
Buck, C.~E., Cavanagh, W.~G., and Litton, C.~D. (1996).
\newblock {\em {The Bayesian Approach to Interpreting Archaeological Data}}.
\newblock Wiley, Chichester.

\bibitem[Buck and Christen, 1998]{Buck.Christen-1998}
Buck, C.~E. and Christen, J.~A. (1998).
\newblock {A novel approach to selecting samples for radiocarbon dating}.
\newblock {\em Journal of Archaeological Science}, 25:303--310.

\bibitem[Buck et~al., 1999]{BCal}
Buck, C.~E., Christen, J.~A., and James, G.~N. (1999).
\newblock {BCal: an on-line Bayesian radiocarbon calibration tool}.
\newblock {\em Internet Archaeology}, 7.
\newblock (http://intarch.ac.uk/journal/issue7/buck/).

\bibitem[Christen, 1994]{Christen-1994}
Christen, J.~A. (1994).
\newblock Summarizing a set of radiocarbon determinations: a robust approach.
\newblock {\em Applied Statistics}, 43(3):489--503.

\bibitem[Christen and Buck, 1998]{Christen.Buck-1998}
Christen, J.~A. and Buck, C.~E. (1998).
\newblock {Sample selection in radiocarbon dating}.
\newblock {\em Applied Statistics}, 47:543--557.

\bibitem[Christen et~al., 1995]{Christen.Clymo.Litton-1995}
Christen, J.~A., Clymo, R.~S., and Litton, C.~D. (1995).
\newblock {A Bayesian approach to the use of $^{14}$C dates in the estimation
  of the age of peat}.
\newblock {\em Radiocarbon}, 37(2):431--442.

\bibitem[Dye and Buck, 2015]{Dye.Buck-2015}
Dye, T.~S. and Buck, C.~E. (2015).
\newblock {Archaeological sequence diagrams and Bayesian chronological models}.
\newblock {\em Journal of Archaeological Science}, 83:84--93.

\bibitem[Haslett and Parnell, 2008]{Haslett.Parnell-2008}
Haslett, J. and Parnell, A. (2008).
\newblock A simple monotone process with application to radiocarbon-dated depth
  chronologies.
\newblock {\em Applied Statistics}, 57:399--418.

\bibitem[Jones, 2013]{Jones-2013}
Jones, E. (2013).
\newblock {\em {Practical Bayesian Dendrochronology}}.
\newblock PhD thesis, University of Sheffield.

\bibitem[Lee and Bronk~Ramsey, 2012]{Lee.BR-2012}
Lee, S. and Bronk~Ramsey, C. (2012).
\newblock Development and application of the trapezoidal model for
  archaeological chronologies.
\newblock {\em Radiocarbon}, 54(1):107--122.

\bibitem[Litton and Zainodin, 1991]{Litton.Zainodin-1991}
Litton, C.~D. and Zainodin, H.~J. (1991).
\newblock Statistical models of dendrochronology.
\newblock {\em Journal of Archaeological Science}, 18:429--440.

\bibitem[Millard, 2002]{Millard-2002}
Millard, A. (2002).
\newblock Bayesian approach to sapwood estimates and felling dates in
  dendrochronology.
\newblock {\em Archaeometry}, 44(1):137--143.

\bibitem[Millard, 2006]{Millard-2006}
Millard, A. (2006).
\newblock {Bayesian analysis of ESR dates, with application to Border Cave}.
\newblock {\em Quaternary Geochronology}, 1(2):159--166.

\bibitem[Naylor and Smith, 1988]{Naylor.Smith-1988}
Naylor, J.~C. and Smith, A. F.~M. (1988).
\newblock An archaeological inference problem.
\newblock {\em Journal of the American Statistical Association},
  83(403):588--595.

\bibitem[Reimer et~al., 2013]{Reimer.etal-2013}
Reimer, P.~J., Bard, E., Bayliss, A., Beck, J.~W., Blackwell, P.~G.,
  Bronk~Ramsey, C., Buck, C.~E., Edwards, R.~L., Friedrich, M., Grootes, P.~M.,
  Guilderson, T.~P., Haflidason, H., Hajdas, I., Hatt\'e, C., Heaton, T.~J.,
  Hoffmann, D.~L., Hogg, A.~G., Hughen, K.~A., Kaiser, K.~F., Kromer, B.,
  Manning, S.~W., Niu, M., Reimer, R.~W., Richards, D.~A., Scott, E.~M.,
  Southon, J.~R., Staff, R.~A., Turney, C. S.~M., and van~der Plicht, J.
  (2013).
\newblock {IntCal13 and Marine13 radiocarbon age calibration curves 0--50,000
  years cal BP}.
\newblock {\em Radiocarbon}, 55(4):1869--1887.

\bibitem[Zink, 2015]{Zink-2015}
Zink, A. (2015).
\newblock Bayesian analysis of luminescence measurements.
\newblock {\em Radiation Measurements}, 81:71--77.

\end{thebibliography}
\bibliographystyle{apalike}

\end{document}